\documentclass[letterpaper]{article}
\usepackage[preprint]{aaai2027}
\usepackage[hyphens]{url}
\usepackage{graphicx}
\usepackage{natbib}
\usepackage{caption}
\usepackage{algorithm}
\usepackage{algorithmic}
\usepackage{multirow} 
\usepackage{newfloat}
\usepackage{listings}
\DeclareCaptionStyle{ruled}{labelfont=normalfont,labelsep=colon,strut=off}
\floatstyle{ruled}
\newfloat{listing}{tb}{lst}{}
\floatname{listing}{Listing}
\usepackage{booktabs}
\usepackage{amsmath}
\usepackage{textcomp}

\title{PRIME: Pre-Retrieval Index Modeling \& Enhancement for Structured Document Retrieval}

\author{
    Cheuk Hei Chung \textsuperscript{}\thanks{Email: chchung634@connect.hkust-gz.edu.cn}
    Li Liu\textsuperscript{}\thanks{Corresponding author: Li Liu (avrillliu@hkust-gz.edu.cn)}
}

\affiliations{
    The Hong Kong University of Science and Technology (Guangzhou)
}

\date{}

\usepackage{amsfonts}
\usepackage{amssymb}
\begin{document}
\maketitle
\begin{abstract}

High-precision retrieval over external knowledge bases is essential for LLM agents to operate effectively. However, existing Information Retrieval (IR) research primarily optimizes query-side retrievers (e.g., ColBERT, BGE) and generators (e.g., Self-RAG, GraphRAG), while reducing corpus-side representation to chunking. This leaves index-side quality an underexplored lever. We propose Pre-Retrieval Index Modeling \& Enhancement (PRIME), an index-side optimization framework built on two complementary tracks: an LLM-driven Semantic Restructuring (SR) track that transforms long documents into structured, information-dense sections, and a deterministic Metadata Synthesis (MS) and Hierarchical Retrieval (HR) track that extracts semantic anchors and heading structure for coarse-to-fine section retrieval. Our evaluations show that deterministic metadata matches structure-aware chunking's retrieval quality while cutting embedding computation by 45.8× and index vector count by 22.4×. On top of these gains, Semantic Restructuring (SR) further improves Answer-F1 by 13.7\% in full deployment, with characterized quality–fidelity–cost trade-offs. Across QASPER, BRIGHT, and NFCorpus, we further characterize three qualitative regimes that determine whether index-side optimization helps or degrades retrieval, while showing that these index-side effects are neutralized under a strong reranker.
\end{abstract}
\section{Introduction}

Efficient information retrieval (IR) is essential for LLM agents, as the quality of retrieved contexts serves as the foundational bottleneck for downstream generation \citep{guu2020realm,lewis2020retrieval, borgeaud2022improving, izacard2023atlas}. While query-side retrievers and generators have been extensively refined—from sparse models \citep{robertson2009probabilistic} to dense dual-encoders \citep{karpukhin2020dense}, late-interaction frameworks \citep{khattab2020colbert}, and cross-encoder rerankers—corpus-side representation has largely remained a fixed input. Conventional preprocessing merely alters text without enhancing its information density or logical structure.

\begin{figure}[htbp]
\centering
\includegraphics[width=0.91\linewidth]{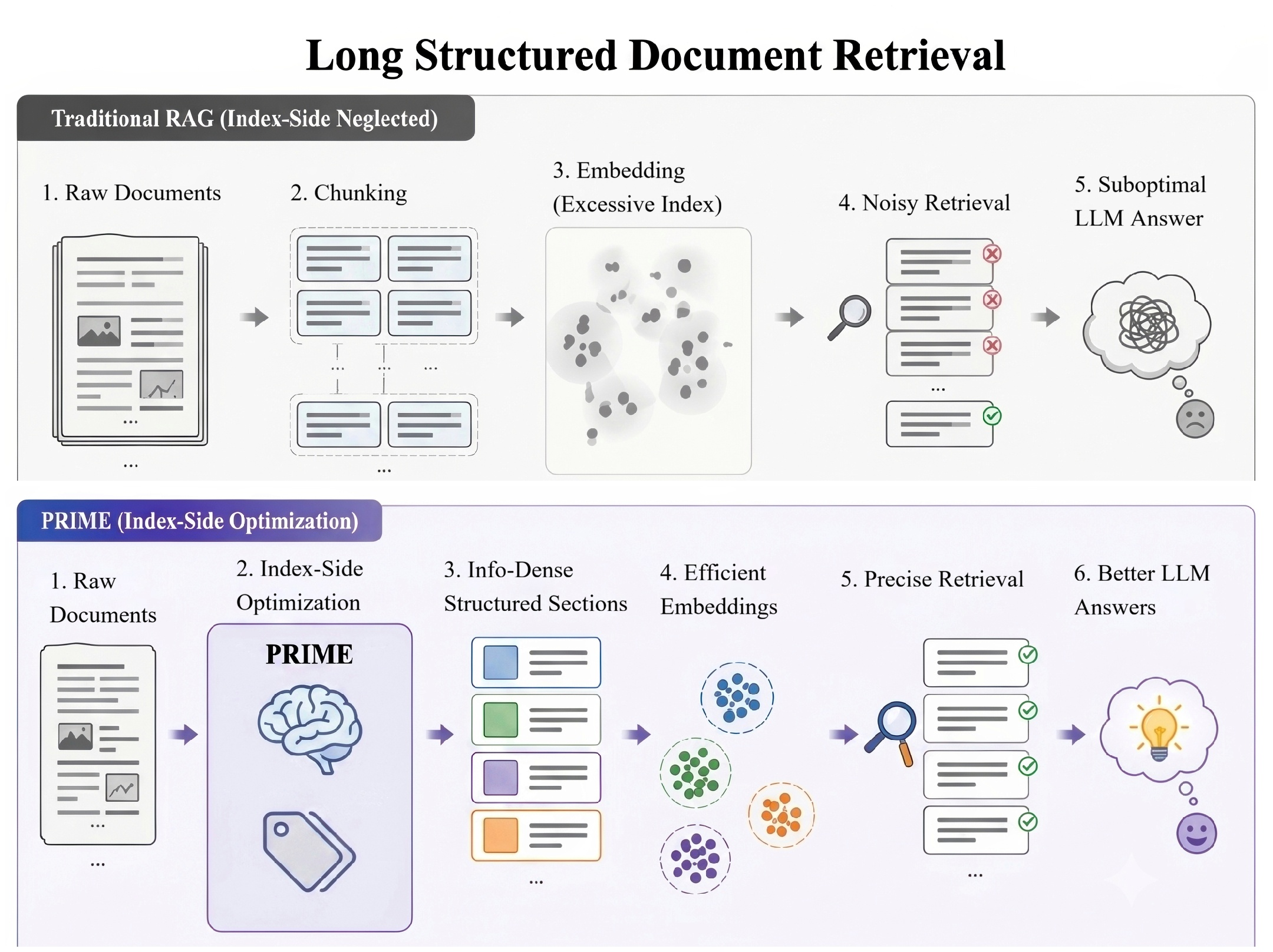}
\caption{Overview of Index-Side Optimization. PRIME operates on the corpus representation prior to indexing, complementing existing query-side and generator-side methods without modifying the retriever.}
\end{figure}

\begin{figure*}[!htbp]
\centering
\includegraphics[width=0.83\textwidth]{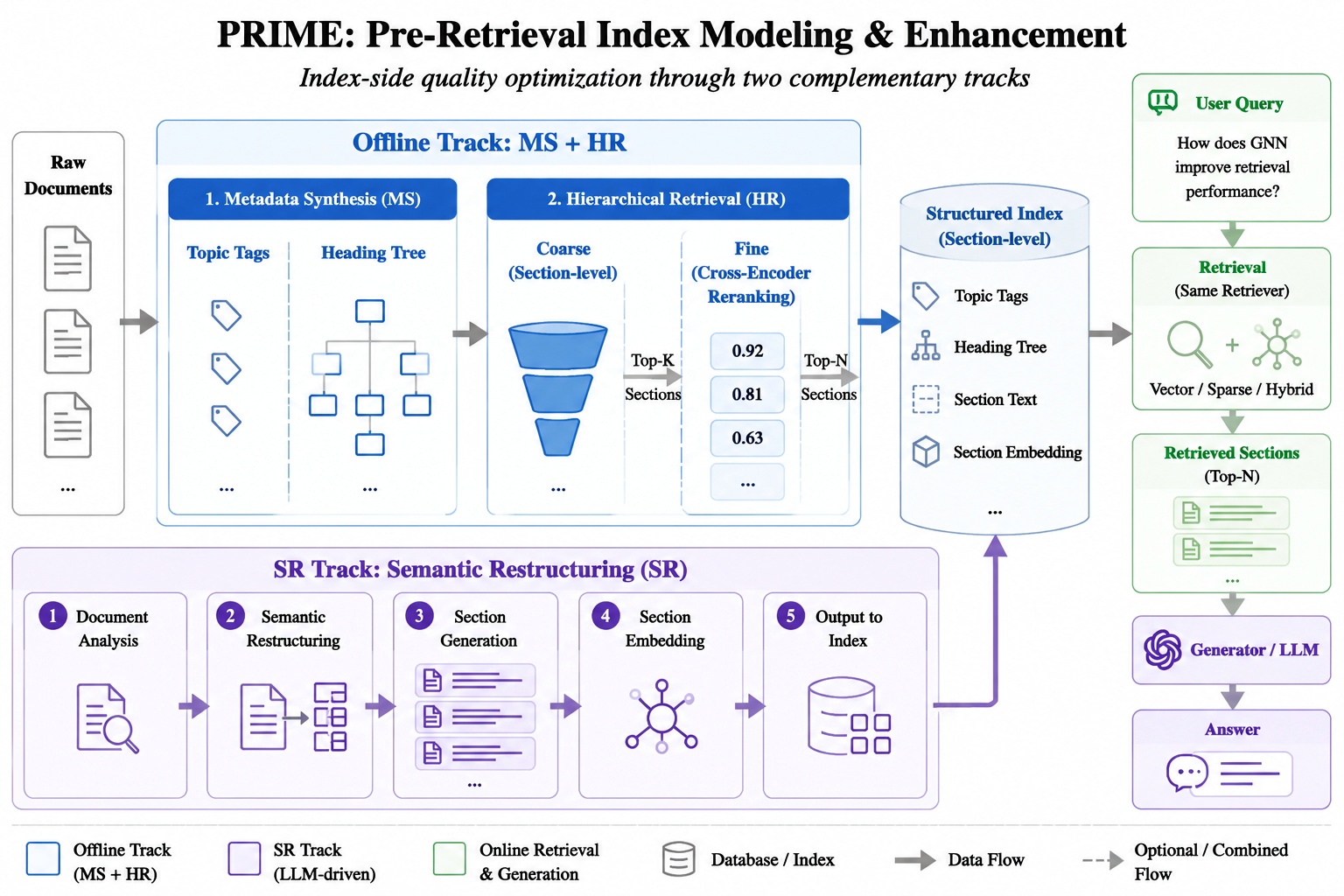}
\caption{PRIME architecture. The framework operates on two complementary tracks: (Left) an LLM-driven Semantic Restructuring track transforms documents into knowledge-dense representations; (Right) a deterministic track uses Metadata Synthesis to extract structural anchors, which power a Hierarchical Retrieval pipeline with RRF fusion and cross-encoder reranking.}
\end{figure*}

Recent methods began to exploit document structure for indexing, and \citet{singh2026beyond} show that once a strong cross-encoder reranker is present, common retrieval enhancements yield no reliable additional gain. This raises the question of whether index-side representation can be improved and whether its benefits are compensated by a strong pipeline.

We propose that retrieval quality is bounded by the information density of the indexed corpus. Once a strong retriever and reranker are in place, the remaining leverage lies in information structure and density at the index level. Guided by this principle, we present PRIME, an index-side quality optimization framework that operates on two complementary tracks without modifying the retriever. Semantic Restructuring (SR) employs an LLM-driven transformation to explicate implicit argument structure and compress descriptive prose within each section, increasing knowledge density for long structured documents. In parallel, Metadata Synthesis (MS) and Hierarchical Retrieval (HR) form a fully deterministic track: MS extracts semantic anchors and heading structure from the original document, which HR then consumes with lightweight high-density embeddings for coarse document filtering and a cross-encoder for fine-grained section reranking. Figure 1 situates PRIME in the retrieval landscape.

Our key contributions are:

\begin{itemize}
  \item Deterministic structural metadata matches structure-aware chunking's retrieval quality while cutting embedding computation by 45.8× and index vector count by 22.4×, establishing a more efficient and lightweight alternative to LLM-mediated indexing.
  \item On top of these deterministic gains, Semantic Restructuring (SR) further yields an additional +13.7\% Answer-F1 in full deployment, with characterized semantic quality, fidelity, and cost trade-offs.
  \item Through a 5 $\times$ 3 factorial ablation on QASPER, BRIGHT, and NFCorpus, we identify three corpus regimes,deficit-bridging, structural exploitation, and structural saturation,that determine whether index-side optimization helps or degrades retrieval, and show that a strong cross-encoder reranker neutralizes these gains.
\end{itemize}

\section{Related Works}

Query-side retrieval has been extensively optimized. From sparse retrieval \citep{robertson2009probabilistic} to learned sparse representations such as SPLADE \citep{formal2021splade}. Dense retrieval progressed through dual-encoder architectures including DPR \citep{karpukhin2020dense}, ANCE \citep{xiong2021approximate} and Contriever \citep{izacard2021unsupervised}, multi-vector late-interaction models represented by ColBERT \citep{khattab2020colbert}, and multilingual general-purpose encoders like BGE-M3 \citep{xiao2024c}. Multi-stage refinement adds cross-encoder rerankers \citep{nogueira2019passage} and LLM-based listwise ranking approaches such as RankGPT \citep{sun2023chatgpt}. On the query formulation side, methods including HyDE \citep{gao2023precise} expand queries via hypothetical passages. However, these methods optimize how queries are encoded, matched, and ranked against a fixed document representation.  

On Corpus-Side, several lines of work have modified the corpus before indexing. Early document preprocessing techniques \citep{manning2008introduction} adjust surface token forms. Early structured document retrieval \citep{wilkinson1994effective} exploited pre-existing field markup for weighted term scoring but relied on author-provided structure. Extractive summarization for IR \citep{tombros1998advantages} selects a subset of existing sentences and trades recall for conciseness without generating new, denser text. Document expansion via query prediction such as Doc2Query \citep{nogueira2019document}, appends synthetic queries to broaden lexical coverage for term matching, yet leaves the document's original knowledge content and information density unchanged. Chunking strategies \citep{bhat2025rethinking} determine where to place segment boundaries but preserve the token composition within each segment. Across these approaches, a common pattern emerges: each modifies how text is packaged, not what the text conveys.  

\textbf{Index-Side Restructuring: Recent Progress and Remaining Gap.} Recent methods have begun exploiting document structure for indexing. Multi-view Content-aware Indexing \citep{dong2024multi}, DISRetrieval \citep{chen2026beyond}, and PaperRegister \citep{wang2026paperregister} organize documents by content structure, discourse, or register but derive representations from original tokens without transformation. Summary-Augmented Chunking \citep{reuter2025towards} injects document-level context into chunks, improving retrieval without rewriting the indexed text. RAPTOR \citep{sarthi2024raptor} generates recursive LLM summaries but relies on embedding-driven clustering. GraphRAG \citep{edge2024from} and follow-up graph-based methods such as PathRAG \citep{chen2026pathrag} and LeanRAG \citep{zhang2026leanrag} reorganize at the graph level without modifying indexed text. Across these methods, structure is exploited but text is not transformed; text is enriched but its information density is not increased. \citet{singh2026beyond} add a critical constraint: once a strong cross-encoder reranker is present, retrieval enhancements yield no Holm-corrected gain. PRIME occupies the resulting gap; generative text restructuring that increases information density (SR), combined with deterministic metadata extraction (MS) and structure-aware hierarchical retrieval (HR), with characterized boundary conditions.

\section{Method}

\subsection{Preliminary}

Let $\mathcal{C} = \{D_1, \ldots, D_N\}$ denote the corpus where each document $D_i$ consists of second-level headings (H2). PRIME processes documents through two tracks, illustrated in Figure 2. \textbf{Semantic Restructuring} (SR) employs LLM-driven analysis to transform $D_i$ into a knowledge-dense representation $D_i^*$ that preserves the original H2 structure. \textbf{Metadata Synthesis (MS) and Hierarchical Retrieval (HR)} form a deterministic track where MS extracts top-5 discriminative terms $\textbf{t}$ and a heading tree $\textbf{h}$ from the corpus ($D_i$ or $D_i^*$), and HR consumes these metadata through a two-stage pipeline: coarse document filtering on high-density metadata embeddings followed by section-level cross-encoder reranking.

PRIME supports two deployment modes. \textbf{Offline mode} runs MS and HR directly on the original corpus without SR, producing metadata $M_i = \text{MS}(D_i)$. \textbf{Full mode} applies SR to produce $D_i^*$, then extracts metadata from the restructured text ($M_i^* = \text{MS}(D_i^*)$) and runs HR; $D_i^*$ also replaces $D_i$ as input to BM25 and the cross-encoder reranker. For component ablation, we also evaluate the SR body text by itself without MS metadata (denoted SR (full-text) in Table 2).

Formally, the two modes are:

\begin{equation}
\begin{aligned}
\text{Offline:} &\quad D_i \xrightarrow{\text{MS}} M_i \\
&\quad I_{\text{vec}} = \text{Index}(M_i) \\
&\quad I_{\text{bm25}} = \text{Index}(D_i) \\[1em]
\text{Full:} &\quad D_i \xrightarrow{\text{SR}} D_i^* \\
&\quad D_i^* \xrightarrow{\text{MS}} M_i^* \\
&\quad I_{\text{vec}} = \text{Index}(M_i^*) \\
&\quad I_{\text{bm25}} = \text{Index}(D_i^*)
\end{aligned}
\end{equation}

where $M_i = (\textbf{t}_i, \textbf{h}_i)$ denotes the top-5 discriminative terms and heading tree.

\subsection{Semantic Document Restructuring}
SR transforms each document $D_i$ into a knowledge-dense representation $D_i^*$ while preserving the original H2 structure. It operates along two semantic axes: \textbf{Explication} reconstructs implicit argumentative structures within each H2 section into declarative statements, and \textbf{Compression} eliminates descriptive prose while retaining core contributions, conclusions, key terms, and numerical results. By removing low-IDF filler tokens, SR shifts $D_i^*$ toward high-IDF discriminative terms, so the cross-encoder's pairwise attention operates on a denser semantic signal per token and the LLM generator consumes fewer tokens for the same answer-relevant content.

To ensure complete coverage, SR invokes the LLM per H2 section following the document's heading tree, mitigating lost-in-the-middle effects \citep{liu2024lost}. The computational cost is $c(D) = \sum_{h \in H2(D)} (\alpha \cdot \text{len}(h) + \beta)$, where $\alpha$ captures per-token inference cost and $\beta$ the per-section prompt and API overhead; output tokens add a symmetric cost factor (full break-even analysis in the Supplementary Material, Appendix D.6).

\subsection{Deterministic Metadata Synthesis}

The Metadata Synthesis (MS) module generates two types of metadata for the HR pipeline (Hierarchical Retrieval): discriminative terms for document-level separation and a heading tree for section-level navigation.

Discriminative terms are selected as the top-5 candidates using a TF-IDF scoring function \citep{salton1988termweighting}:

\begin{multline}
\text{score}(t)
  = \text{TF}(t) \times \text{IDF}_{\text{corpus}}(t) \\
  \times (1-\alpha)^{\text{first\_pos}/n}
    \times \bigl(1 + \beta \cdot \mathbb{1}_{\text{in\_heading}}(t)\bigr).
\end{multline}

where $\text{TF}(t)$ and $\text{IDF}_{\text{corpus}}(t)$ denote the term frequency and corpus-level inverse document frequency of $t$, $\mathrm{first\_pos}$ is the token position of $t$'s first occurrence in $D_i$, $n = |D_i|$, $\alpha = 0.3$ controls the positional decay rate, and $\beta = 2.0$ weights the heading bonus. Terms are whitespace-delimited tokens after lowercasing and punctuation removal. $\text{IDF}_{\text{corpus}}$ uses the smoothed form $\log\frac{N+1}{\text{df}(t)+1}+1$ to avoid negative values when $\text{df}(t)=N$, and is computed from the evaluation corpus itself rather than from an external reference, since domain-specific IDF better captures distinctiveness across structurally diverse corpora (evaluated in Experiments).

The heading tree, parsed from H2 headings of $D_i$, together with the document title and discriminative terms form a high-density metadata embedding for coarse filtering and provide section boundaries for fine-grained reranking.

\subsection{Structure-Aware Hierarchical Retrieval}

The Hierarchical Retrieval (HR) module consumes MS-generated metadata through a two-stage pipeline: coarse document filtering followed by section-level reranking. This design reflects the observation that document-level relevance can be assessed via coarse semantic signals, while section-level relevance requires fine-grained token interaction.

\subsubsection{Stage 1 — High-Density Document Filtering}

Stage 1 retrieves candidate documents using two parallel pathways. The BM25 pathway indexes the full document text for exact keyword matches. The vector pathway indexes the high-density metadata (title + discriminative terms + heading tree) encoded via BGE-small-en-v1.5 (384-dim; Xiao et al., 2024). Results from both pathways are merged via Reciprocal Rank Fusion (RRF, $k = 60$) \citep{cormack2009reciprocal} to form a top-30 candidate set. The heading tree is invariant to whether SR is applied (H2 structure is preserved). In PRIME (offline), discriminative terms are extracted from $D_i$. In full PRIME, terms are extracted from $D_i^*$, reflecting SR's vocabulary shift; the resulting impact is quantified in the Ablation.

\subsubsection{Stage 2 — Section-Level Evidence Localization}

Stage 2 scores each H2 section of candidate documents using a cross-encoder (ms-marco-MiniLM-L-6-v2; Reimers \& Gurevych, 2019), fine-tuned on MS MARCO \citep{bajaj2016msmarco}. The top-5 sections by CE score are forwarded to the LLM generator. Document-level rankings in Experiments are derived via best-section aggregation: $\text{score}(D) = \max_{s \in D} \text{CE}(q, s)$. The lightweight cross-encoder eliminates query-time LLM overhead typical of listwise rerankers. Full model configuration and hyperparameters are provided in Appendix B.

\section{Experiments}

\subsubsection{Datasets}

We evaluate on three structurally diverse benchmarks: \textbf{QASPER} \citep{dasigi2021dataset} serves as the target scenario for PRIME's full pipeline: we use 416 papers from the official test set, producing 1,451 queries with document-level qrels (one relevant document per query, 415 negatives). Queries are natural language questions without title or metadata. \textbf{BRIGHT} \citep{su2025bright} tests cross-domain generalization of PRIME's offline mode across mixed-structure sources. We evaluate all twelve subsets despite four (AoPS, LeetCode, TheoremQA-Q, and TheoremQA-Theorems) requiring reasoning which fall outside PRIME's design scope. \textbf{NFCorpus} \citep{boteva2016fulltext}, from the BEIR benchmark \citep{thakur2021beir}, defines the boundary condition where structural metadata provides no discriminative signal.

\subsubsection{Evaluation Metrics}

We evaluate PRIME across four complementary dimensions: retrieval quality, downstream QA utility, text density, and operational cost. This evaluation framework establishes a complete evidence chain from index-side representation quality to end-to-end task performance.

\textbf{Retrieval Quality}. All three datasets use NDCG@5, P@5, R@5, P@1, R@1, and MRR. QASPER uses document-level qrels (one relevant doc per query), so P@1 $\equiv $ R@1 and R@5 = 5×P@5, and NDCG@5 reduces to a rank-based metric driven entirely by the position of the single relevant document. NFCorpus uses graded relevance (0–2); BRIGHT uses evidence-based binary qrels.

\textbf{Consumption-Side QA Quality.} QA is evaluated via QASPER's official Answer-F1 \citep{dasigi2021dataset}, computed as the best match over the three reference answers per question.

\textbf{Text Density — Corpus-Level Transformation.} We quantify SR's effect on document representation via four metrics. Three capture orthogonal dimensions: \textbf{stopword ratio} (filler compression), \textbf{technical term density} (domain terms per 100 tokens), and \textbf{mean IDF} (average token distinctiveness) \citep{robertson2004understanding}. The fourth, \textbf{semantic density}, integrates them into a composite signal-to-noise ratio. We define semantic density as:

\[\text{Semantic Density} = \frac{\text{content\_tokens} \times \text{mean\_idf}}{\text{stop\_tokens} + 1}\]

The numerator estimates total information carried by a document, content-bearing tokens weighted by their average distinctiveness across the corpus. The denominator estimates noise, filler words that contribute surface length without semantic content. A full derivation appears in Appendix D.7.

\subsubsection{Implementation}

All PRIME variants share a unified backend: BGE-small-en-v1.5 (384-dim) \cite{xiao2024c}, hnswlib configured with M=32, ef=200, and cosine similarity \citep{malkov2018efficient}, and SQLite FTS5 with porter stemming. All query-side components (embedding, BM25, vector search, and cross-encoder (ms-marco-MiniLM-L-6-v2) \citep{reimers2019sentence}) run on a single NVIDIA RTX 5090 GPU. SR uses DeepSeek-V4-pro \citep{deepseek2026deepseekv4} at temp=0; QA generation in Context Quality uses DeepSeek-V4-Flash. All hyperparameters were fixed before evaluation based on literature defaults with no tuning on test queries. Full configuration and hyperparameter tables are in Appendix D.

\subsection{Main Results}
\subsubsection{Retrieval Quality}

\paragraph{QASPER}

Index-side optimization improves retrieval, but its measurable advantage narrows as downstream pipeline components increasingly compensate for weaker embeddings. We establish this through two complementary analyses: a shared-pipeline ceiling that reveals convergence under a strong reranker (Table 1), and a factorial decomposition that isolates index representation from pipeline effects (Table 2).

\begin{table*}[!htbp]
\centering
\footnotesize
\caption{Retrieval quality on QASPER 416 (1,451 queries). \textit{BGE-M3 uses chunk-256 max-pooling; Summary-Augmented prepends section summaries to chunks; RAPTOR-lite does section+window retrieval without LLM summarization. Structure-chunks is a simpler H2-aware chunking baseline (no MS metadata, no HR pipeline). PRIME (Full) is excluded because SR alters the BM25-indexed text, breaking the shared-pipeline condition.}}
\begin{tabular}{lcccccc}
\hline
Method & NDCG\@5 & NDCG\@10 & Recall\@5 & Precision\@5 & P\@1 & MRR \\
\hline
Raw (chunk-256) & 0.391 & 0.405 & 0.462 & 0.092 & 0.306 & 0.373 \\
BGE-M3 (1024-dim) & 0.389 & 0.398 & 0.453 & 0.091 & 0.306 & 0.371 \\
Summary-Augmented & 0.398 & 0.411 & 0.471 & 0.094 & 0.309 & 0.379 \\
RAPTOR-lite & 0.391 & 0.408 & 0.461 & 0.092 & 0.307 & 0.375 \\
PRIME (offline) \textbf{(ours)} & 0.\textbf{396} & 0.409 & \textbf{0.469} & \textbf{0.094} & 0.309 & 0.377 \\
Structure-chunks & \textbf{0.396} & \textbf{0.411} & 0.466 & 0.093 & \textbf{0.310} & \textbf{0.379} \\
\hline
\end{tabular}
\end{table*}

All six methods converge within NDCG@5 = 0.389–0.398, echoing the reranker-saturation effect reported by \citet{singh2026beyond}: once a strong cross-encoder is in the loop, corpus-side differences become nearly inseparable. This apparent equivalence conflates index quality with pipeline strength. We therefore evaluate a $5 \times 3$ factorial matrix varying only the indexed text representation across three pipelines (Dense-only, Hybrid BM25+RRF, Hybrid+CE), disentangling the two. 95\% paired bootstrap CIs (10,000 resamples) are reported.

\begin{table*}[!htbp]
\centering
\footnotesize
\caption{Factorial ablation: NDCG@5 on QASPER 416. \textit{95\% paired bootstrap CIs in brackets. All configurations share identical retrieval pipelines; only the indexed text varies between rows. The five index representations and three retrieval pipelines are defined in Preliminary.}}
\begin{tabular}{l c c c}
\hline
Index Representation & Dense-only & Hybrid (BM25+Vec+RRF) & Hybrid+CE \\
\hline
Raw (chunk-256) & 0.175 [.164,.186] & 0.311 [.296,.327] & 0.391 [.370,.412] \\
Structure-chunks & 0.201 [.187,.216] & 0.338 [.321,.354] & 0.396 [.375,.416] \\
PRIME (offline) & 0.201 [.186,.217] & 0.332 [.315,.349] & 0.396 [.375,.416] \\
SR (full-text) & 0.208 [.196,.221] & 0.245 [.232,.258] & 0.328 [.309,.347] \\
PRIME (Full) & 0.214 [.199,.229] & 0.241 [.227,.256] & 0.325 [.307,.345] \\
\hline
\end{tabular}
\end{table*}

As shown in Table 2, \textbf{Index representation matters in isolation}: in the Dense-only column, all four structured representations outperform chunk-256 ($p<0.001$). The convergence in table 1 reflects that BM25 and the CE reranker mask these underlying embedding differences: the advantage of structured over raw narrows from +15–22\% (Dense-only) to +7–9\% (Hybrid) to <0.005 (Hybrid+CE).

Within this narrowing gap, structured representations exhibit distinct efficiency trade-offs: \textbf{Deterministic metadata matches structure-aware chunking at a fraction of the cost}. Structure-chunks and PRIME (offline) remain indistinguishable (overlapping 95\% CIs). Yet, PRIME (offline) embeds only metadata while capturing the same document-filtering signal as full-text chunking, at 45.8× less embedding computation and 22.4× fewer indexed vectors (Operational Cost). In contrast, both SR-based representations \textbf{trade lexical signal for semantic density}. PRIME (Full) achieves the highest Dense-only NDCG@5 (0.214), confirming that the compact metadata embedding (title, topic tags, and heading tree extracted from SR text, \textasciitilde{}200 tokens) already captures stronger discriminative signal than raw chunk-256. However, BM25—common to both SR (full-text) and PRIME (Full) via the same SR body index—drops substantially below Raw in the Hybrid pipeline (0.241 and 0.245 vs. 0.311), because rewriting shifts text toward high-IDF terms at the expense of surface keyword overlap. CE reranking partially recovers this loss (Hybrid+CE: 0.325 and 0.328), though this retrieval-side cost is ultimately offset by improved context quality during generation (Context Quality).

\paragraph{BRIGHT}
We evaluate PRIME (offline) on all twelve BRIGHT tasks. Table 3a reports the eight retrieval tasks; TheoremQA-Theorems is excluded from the average and noted in the caption. The other three reasoning tasks are reported separately in Appendix E of Supplementary Material.

\begin{table*}[htbp]
  \centering
  \footnotesize
  \caption{Cross-domain retrieval on the eight BRIGHT structured tasks. \textit{$\Delta $ = (PRIME-Baseline)/Baseline; 95\% CIs from 10,000-sample paired bootstrap (BCa) over per-query NDCG@5. Per-task CIs span zero (65-103 queries/task), so individual task deltas are not independently significant; the +6.9\% macro-average reflects a distributional shift across domains rather than a per-task claim. TheoremQA-Theorems (PRIME NDCG@5 = 0.012, $\Delta $ = -59.8\%) is excluded from this average.}}
  \label{tab:prime-results}
  \begin{tabular}{lcccccccl}
    \toprule
    \multirow{2}{*}{Tasks} 
      & \multicolumn{3}{c}{BGE-small (baseline)} 
      & \multicolumn{3}{c}{PRIME (offline)} 
      & \multirow{2}{*}{$\Delta$ NDCG@5} 
      & \multirow{2}{*}{95\% CI ($\Delta$)} \\
    \cmidrule(lr){2-4} \cmidrule(lr){5-7}
      & NDCG@5 & P@5 & R@5 
      & NDCG@5 & P@5 & R@5 
      & & \\
    \midrule
    Pony              
      & 0.067 & 0.073 & 0.048 
      & 0.185 & 0.154 & 0.126 
      & $+175.6\%$ & $[-0.063,\ +0.299]$ \\
    Biology           
      & 0.379 & 0.128 & 0.491 
      & 0.464 & 0.155 & 0.598 
      & $+22.3\%$  & $[-0.188,\ +0.357]$ \\
    Stackoverflow     
      & 0.234 & 0.075 & 0.359 
      & 0.272 & 0.079 & 0.363 
      & $+16.2\%$  & $[-0.203,\ +0.279]$ \\
    Robotics          
      & 0.181 & 0.052 & 0.253 
      & 0.200 & 0.059 & 0.282 
      & $+10.5\%$  & $[-0.250,\ +0.290]$ \\
    Economics         
      & 0.247 & 0.070 & 0.333 
      & 0.258 & 0.074 & 0.354 
      & $+4.6\%$   & $[-0.242,\ +0.265]$ \\
    Sustainable\_living
      & 0.250 & 0.080 & 0.342 
      & 0.242 & 0.082 & 0.343 
      & $-3.1\%$   & $[-0.259,\ +0.244]$ \\
    Psychology        
      & 0.217 & 0.069 & 0.304 
      & 0.199 & 0.061 & 0.259 
      & $-8.2\%$   & $[-0.277,\ +0.242]$ \\
    Earth\_science    
      & 0.458 & 0.157 & 0.537 
      & 0.352 & 0.128 & 0.419 
      & $-23.1\%$  & $[-0.404,\ +0.193]$ \\
    \midrule
    \textbf{Avg (8 tasks)} 
      & \textbf{0.254} & \textbf{0.088} & \textbf{0.333} 
      & \textbf{0.272} & \textbf{0.099} & \textbf{0.343} 
      & \textbf{$+6.9\%$} & --- \\
    \bottomrule
  \end{tabular}
\end{table*}

The $\Delta $ distribution across the eight structured tasks falls into three qualitative regimes, each predicted by corpus properties that can be assessed before retrieval. At one extreme, \textbf{deficit-bridging} (pony, +175.6\%): dense API reference documentation is poorly represented by BGE-small's pretraining distribution, so the discriminative terms by MS serve as a natural-language bridge for the encoder. Typical structured documents earn modest gains from \textbf{structure-exploiting} (biology to economics, +4.6\% to +22.3\%), where structured H2 sections with \texttt{[edit]} wiki markers anchor CE section reranking. At the opposite end, \textbf{structure-absent or saturated} domains (sustainable\_living to earth\_science, -3.1\% to -23.1\%) show that prose already well-embedded by BGE-small, with sparse or empty heading trees, receives no additional discriminative signal from metadata. TheoremQA-Theorems (-59.8\%, excluded from the average) marks a distinct fourth mode rather than saturation: its LaTeX blocks are parsed as H2 sections but fragment mathematical proofs and requires reasoning, so the CE scores an isolated definition against a query that requires a complete inferential chain.

\paragraph{NFCorpus}

NFCorpus \citep{boteva2016fulltext} defines the opposite boundary. This dataset features short documents without H2 headings, PRIME extracts only a pseudo-title with five TF-IDF topic tags per document. We also report PRIME (Full-Text) where the same topic tags are injected alongside the original full-text embedding to isolate the metadata contribution.

\begin{table}[!htbp]
\centering
\small
\caption{Retrieval quality on the full NFCorpus. Graded relevance 0--2 binarized at $\geq1$. PRIME(FT) denotes PRIME (Full-Text), $\Delta_1$ denotes PRIME vs BGE-small and $\Delta_2$ denotes PRIME (Full-Text) vs BGE-small. }
\begin{tabular}{lcccccc}
\hline
Method & NDCG@5 & P@5 & R@5 &  MRR \\
\hline
BM25 & 0.350 & 0.295 & 0.119 &  0.530 \\
BGE-small & 0.374 & 0.324 & 0.121 &  0.538 \\
BGE-base & 0.387 & 0.336 & 0.126 & 0.561 \\
PRIME & 0.301 & 0.264 & 0.084 &  0.460 \\
PRIME (FT) & 0.366 & 0.316 & 0.118 & 0.540 \\
$\Delta_1$ & -19.5\% & -18.5\% & -30.6\% &  -14.5\% \\
$\Delta_2$  & -2.1\% & -2.5\% & -2.5\% &  +0.4\% \\
\hline
\end{tabular}
\end{table}

As shown in Table 4, PRIME alone drops 19.5\% NDCG@5 below BGE-small: the five topic tags are insufficient context for short, high-density abstracts. In contrast, PRIME (Full-Text) recovers to within 2.1\% of BGE-small (NDCG@5 0.366 vs 0.374), slightly exceeding in MRR, confirming that the tags introduce a mild perturbation without adding discriminative signal, the boundary condition where structural metadata ceases to help and begins to interfere.

\subsubsection{Context Quality}
Retrieval Quality showed that SR degrades retrieval under Hybrid+CE (NDCG@5 0.325 vs. 0.391 for Raw). We find that SR's downstream gains in context quality partially compensate for this retrieval cost. Table 5 traces the causal chain from text density through evidence localization to answer quality.

Table 5 quantifies SR's effect at two intermediate stages. Panel A reveals that SR increases the knowledge density of the context substantially: filler tokens drop by 36.3\%, mean IDF rises 19.7\%, and semantic density increased by 127.3\%. Panel B shows that this signal propagates to all evidence retrieval metrics on the 708 queries with section-level evidence in both corpora: SR+CE outperforms Raw+CE on section-level NDCG@5 (+45.7\%), Recall@5 (+59.9\%), Precision@5 (+7.9\%), MAP (+69.4\%), MRR (+15.4\%), and Token Recall@5 (+14.1\%), while indexing 29.0\% fewer evidence tokens per query, confirming that the CE better localizes relevant content within fewer tokens when the underlying text is denser. Across QASPER's question categories, SR improves section-level Recall@5 for extractive (+63\%), abstractive (+56\%), and unanswerable (+61\%) queries, indicating the gain stems from text density rather than answer format. Yes/no queries are excluded (2 queries with evidence-bearing sections). Full derivation of density metrics in Appendix D.5, full evidence retrieval metrics in Appendix D.7.

\begin{table}[!htbp]
\centering
\footnotesize
\caption{SR causal chain: text density and evidence localization (QASPER 416). Panel A uses fixed raw-corpus reference IDF. Panel B uses section qrels from gold evidence spans mapped to H2 sections, averaged over the 708 shared queries of Raw and SR section qrels to avoid survivor bias.}
\begin{tabular}{lrrr}
\hline
& Raw & SR & $\Delta$ \% \\
\hline
\multicolumn{4}{l}{\textbf{Panel A: Corpus-level density} (416 docs)} \\
\hline
Stopword ratio & 0.365 & 0.233 & -36.3\% \\
Mean IDF & 1.947 & 2.331 & +19.7\% \\
Tech term density (per 100 toks) & 46.3 & 53.8 & +16.1\% \\
Semantic density (composite) & 3.42 & 7.76 & +127.3\% \\
\hline
\multicolumn{4}{l}{\textbf{Panel B: Section-level evidence retrieval} (CE reranker)} \\
\hline
NDCG@5 & 0.127 & 0.185 & +45.7\% \\
Recall@5 & 0.128 & 0.204 & +59.9\% \\
Precision@5 & 0.089 & 0.096 & +7.9\% \\
MAP & 0.095 & 0.161 & +69.4\% \\
MRR & 0.198 & 0.229 & +15.4\% \\
Token Recall@5 & 0.442 & 0.505 & +14.1\% \\
Gold evidence toks per query & 346.9 & 246.3 & -29.0\% \\
\hline
\end{tabular}
\end{table}

Finally, we quantify SR's end-to-end QA impact, denoting configurations as retrieval corpus to generation corpus. All configurations share the same full-body hybrid retrieval pipeline, differing only in whether the raw or SR corpus being indexed and read. Raw to Raw yields a baseline Answer-F1 of 0.0842. Raw to SR adds +2.9\% (0.0866), isolating SR's value as denser reading material. Full to Full reaches 0.0957 (+13.7\%), despite its document-level retrieval being substantially weaker (demonstrated in Table 2). The gap between Raw to SR and Full to Full shows that SR benefits QA primarily through improved cross-encoder evidence localization on denser sections, which more than compensates for the document-level retrieval drop.

\subsubsection{SR Fidelity}
SR increases text density but risks discarding answer-critical content or introducing factual errors; Table 6 summarizes three audit dimensions across evidence retention, entity preservation, and NLI entailment on QASPER (416 documents, 5,589 raw to SR section pairs, 2,231 gold evidence spans).

\begin{table*}[!htbp]
\centering
\footnotesize
\caption{SR fidelity audit on QASPER. Panel A: a span is retained if its character-level content appears verbatim in the SR block. Panel B uses spaCy en\_core\_web\_sm. Panel C uses roberta-large-mnli.}
\begin{tabular}{ll}
Metrics& Value \\
\hline
\textbf{Panel A: Evidence retention} (2,231 gold spans, 373 docs) & \\
Overall span retention & 63.2\% \\
Median per-document retention & 66.7\% (p25: 44.4\%, p75: 100\%) \\
Perfect retention (100\% of spans) & 26.8\% of docs \\
High retention (80--99\%) & 35.1\% of docs \\
Low retention (<50\%) & 25.7\% of docs \\
\hline
\textbf{Panel B: Entity \& signal preservation} (raw$\rightarrow$SR, paired) & \\
Numerical values preserved & 91.0\% \\
Acronyms preserved & 92.5\% \\
Named entities preserved & 72.7\% \\
Extraneous spans (hallucinated) & 1.3\% \\
\hline
\textbf{Panel C: NLI entailment} (roberta-large-mnli, 5,589 pairs) & \\
Entailment & 4,061 (72.7\%) \\
Neutral & 1,286 (23.0\%) \\
Contradiction & 242 (4.3\%) \\
\end{tabular}
\end{table*}

As shown in Table 6, 63.2\% of evidence spans survive restructuring; losses are almost exclusively through compression rather than hallucination (only 1.3\% of spans appear in SR but not raw text). The 25.7\% of documents with retention below 50\% share a recognizable profile: multi-step procedural reasoning or cross-section comparisons whose intermediate steps SR's declarative reformulation flattens into short summaries. Panel B shows strong preservation of atomic signals (numerical values 91.0\% and acronyms 92.5\%); named entity rates drop to 72.7\%, reflecting repeated mentions collapsing under compression. Panel C reveals a 4.3\% contradiction rate (95.7\% NLI-consistent); 181 of 416 documents have at least one contradiction, concentrated in formal/mathematical content where declarative reformulation flattens conditional or quantified statements.

Taken together, these three audits define the quality-fidelity profile of SR. For roughly 60\% of documents, SR delivers substantially denser text at negligible factual cost. For the remaining documents, the primary risk is omission under compression, not fabrication. Therefore, SR is best suited to declarative, evidence-dense documents while domains rich in procedural reasoning or formal logic require caution.

\subsubsection{Operational Cost}

PRIME (offline) consumes zero query-time LLM tokens and matches structure-aware chunking in retrieval quality (Retrieval Quality, overlapping 95\% CIs). SR adds a one-time preprocessing cost of \~56K tokens/doc including per-section prompt overhead (Semantic Restructuring). Against LLM pointwise reranking (\textasciitilde{}1,926 tokens/query per 20-candidate batch), SR breaks even at \textbf{12,096 queries}, roughly 8.3× QASPER's test set, after which it too operates at zero per-query LLM cost. Full resource accounting tables, per-dataset cost breakdowns, and break-even analysis are in Appendix D.6.

\subsection{Ablation}
The $5 \times 3$ factorial design of Table 2 constitutes a component-wise ablation, isolating index representations under fixed retrieval pipelines. Under Dense-only, PRIME (Full) leads at 0.214 (+22.2\% over raw, $p < 0.001$); PRIME (offline) matches structure-chunks (0.201 vs. 0.201). Under Hybrid+CE, SR-based variants drop below raw (0.325–0.328 vs. 0.391, $p < 0.001$), confirming that CE reranking on unmodified H2 sections saturates document-level signal—index-side gains primarily manifest at the context level (Context Quality).

\section{Discussion}

Our results reveal that index-side optimization is effective at the context level. However, the fact that deterministic metadata matches structured chunking while SR's QA gains flow primarily through improved retrieval rather than context substitution points to a broader issue: retrieval metrics alone are an incomplete lens on corpus-side preprocessing. Additionally, the three-regime pattern across BRIGHT points to a broader implication. Index-side interventions are not universally beneficial, their utility is predictable from corpus properties measurable before retrieval. This shifts the evaluation focus from absolute improvement to corpus-conditioned utility, precisely the dimension that standard IR benchmarks, averaged over heterogeneous corpora, cannot resolve. Four open problems follow. SR's quality-fidelity trade-off raises the question of when rewriting can be trusted without verification; its break-even cost (Operational Cost) motivates lighter restructuring strategies without losing information; the interference on heading-free corpora (Retrieval Quality, NFCorpus) suggests corpus-adaptive strategies; and our text-only pipeline is inherently limited to textual evidence — table and figure content in QASPER is not indexed. These call for a principled theory of index-side optimization conditioned on document structure, rather than empirical heuristics.

\section{Conclusion}

We presented PRIME, an index-side optimization framework combining LLM-driven Semantic Restructuring with deterministic Metadata Synthesis and Hierarchical Retrieval. Deterministic metadata matches structure-aware chunking at 45.8× less embedding computation and 22.4× fewer indexed vectors. On top of these gains, Semantic Restructuring further improves Answer-F1 by 13.7\% in full deployment, with characterized quality–fidelity–cost trade-offs. Through extensive evaluation, we identified the operating conditions and cost profiles of this framework. Future work includes improving SR fidelity under text restructuring in order to fulfill its retrieval potential.

\bibliography{08_References}

@inproceedings{borgeaud2022improving,
  title={Improving language models by retrieving from trillions of tokens},
  author={Borgeaud, Sebastian and Mensch, Arthur and Hoffmann, Jordan and Cai, Trevor and Rutherford, Eliza and Millican, Katie and Van Den Driessche, George Bm and Lespiau, Jean-Baptiste and Damoc, Bogdan and Clark, Aidan and others},
  booktitle={International conference on machine learning},
  pages={2206--2240},
  year={2022},
  organization={PMLR}
}

@article{bajaj2016msmarco,
  title={MS MARCO: A human generated machine reading comprehension dataset},
  author={Bajaj, Payal and Campos, Daniel and Craswell, Nick and Deng, Li and Gao, Jianfeng and Liu, Xiaodong and Majumder, Rangan and McNamara, Andrew and Mitra, Bhaskar and Nguyen, Tri and others},
  journal={arXiv preprint arXiv:1611.09268},
  year={2016}
}

@inproceedings{boteva2016fulltext,
  title={A full-text learning to rank dataset for medical information retrieval},
  author={Boteva, Vera and Gholipour, Demian and Sokolov, Artem and Riezler, Stefan},
  booktitle={European Conference on Information Retrieval},
  pages={716--722},
  year={2016},
  organization={Springer}
}

@inproceedings{cormack2009reciprocal,
  title={Reciprocal rank fusion outperforms condorcet and individual rank learning methods},
  author={Cormack, Gordon V and Clarke, Charles LA and Buettcher, Stefan},
  booktitle={Proceedings of the 32nd international ACM SIGIR conference on Research and development in information retrieval},
  pages={758--759},
  year={2009}
}

@inproceedings{chen2026pathrag,
  title={PathRAG: Pruning Graph-based Retrieval Augmented Generation with Relational Paths},
  author={Chen, Boyu and Guo, Zirui and Yang, Zidan and Chen, Yuluo and Chen, Junze and Liu, Zhenghao and Shi, Chuan and Yang, Cheng},
  booktitle={Proceedings of the AAAI Conference on Artificial Intelligence},
  volume={40},
  pages={30183--30191},
  year={2026}
}

@inproceedings{xiao2024c,
  title={C-pack: Packed resources for general chinese embeddings},
  author={Xiao, Shitao and Liu, Zheng and Zhang, Peitian and Muennighoff, Niklas and Lian, Defu and Nie, Jian-Yun},
  booktitle={Proceedings of the 47th international ACM SIGIR conference on research and development in information retrieval},
  pages={641--649},
  year={2024}
}

@inproceedings{dasigi2021dataset,
  title={A dataset of information-seeking questions and answers anchored in research papers},
  author={Dasigi, Pradeep and Lo, Kyle and Beltagy, Iz and Cohan, Arman and Smith, Noah A and Gardner, Matt},
  booktitle={Proceedings of the 2021 Conference of the North American Chapter of the Association for Computational Linguistics: Human Language Technologies},
  pages={4599--4610},
  year={2021}
}

@article{deepseek2026deepseekv4,
  title={Deepseek-v4: Towards highly efficient million-token context intelligence},
  author={Xu, Anyi and Lin, Bangcai and Xue, Bing and Wang, Bingxuan and Xu, Bingzheng and Wu, Bochao and Zhang, Bowei and Lin, Chaofan and Dong, Chen and Ling, Chenchen and others},
  journal={arXiv preprint arXiv:2606.19348},
  year={2026}
}

@article{edge2024from,
  title={From local to global: A graph rag approach to query-focused summarization},
  author={Edge, Darren and Trinh, Ha and Cheng, Newman and Bradley, Joshua and Chao, Alex and Mody, Apurva and Truitt, Steven and Metropolitansky, Dasha and Ness, Robert Osazuwa and Larson, Jonathan},
  journal={arXiv preprint arXiv:2404.16130},
  year={2024}
}

@inproceedings{formal2021splade,
  title={Splade: Sparse lexical and expansion model for first stage ranking},
  author={Formal, Thibault and Piwowarski, Benjamin and Clinchant, St{\'e}phane},
  booktitle={Proceedings of the 44th International ACM SIGIR Conference on Research and Development in Information Retrieval},
  pages={2288--2292},
  year={2021}
}

@inproceedings{gao2023precise,
  title={Precise zero-shot dense retrieval without relevance labels},
  author={Gao, Luyu and Ma, Xueguang and Lin, Jimmy and Callan, Jamie},
  booktitle={Proceedings of the 61st Annual Meeting of the Association for Computational Linguistics (Volume 1: Long Papers)},
  pages={1762--1777},
  year={2023}
}

@InProceedings{guu2020realm,
  title = 	 {Retrieval Augmented Language Model Pre-Training},
  author =       {Guu, Kelvin and Lee, Kenton and Tung, Zora and Pasupat, Panupong and Chang, Mingwei},
  booktitle = 	 {Proceedings of the 37th International Conference on Machine Learning},
  pages = 	 {3929--3938},
  year = 	 {2020},
  editor = 	 {III, Hal Daumé and Singh, Aarti},
  volume = 	 {119},
  series = 	 {Proceedings of Machine Learning Research},
  month = 	 {13--18 Jul},
  publisher =    {PMLR},
  url = 	 {https://proceedings.mlr.press/v119/guu20a.html},
}

@article{izacard2021unsupervised,
  title={Unsupervised dense information retrieval with contrastive learning},
  author={Izacard, Gautier and Caron, Mathilde and Hosseini, Lucas and Riedel, Sebastian and Bojanowski, Piotr and Joulin, Armand and Grave, Edouard},
  journal={arXiv preprint arXiv:2112.09118},
  year={2021}
}

@article{izacard2023atlas,
  title={Atlas: Few-shot learning with retrieval augmented language models},
  author={Izacard, Gautier and Lewis, Patrick and Lomeli, Maria and Hosseini, Lucas and Petroni, Fabio and Schick, Timo and Dwivedi-Yu, Jane and Joulin, Armand and Riedel, Sebastian and Grave, Edouard},
  journal={Journal of Machine Learning Research},
  volume={24},
  number={251},
  pages={1--43},
  year={2023}
}

@inproceedings{karpukhin2020dense,
  title={Dense passage retrieval for open-domain question answering},
  author={Karpukhin, Vladimir and Oguz, Barlas and Min, Sewon and Lewis, Patrick and Wu, Ledell and Edunov, Sergey and Chen, Danqi and Yih, Wen-tau},
  booktitle={Proceedings of the 2020 conference on empirical methods in natural language processing (EMNLP)},
  pages={6769--6781},
  year={2020}
}

@inproceedings{khattab2020colbert,
  title={Colbert: Efficient and effective passage search via contextualized late interaction over bert},
  author={Khattab, Omar and Zaharia, Matei},
  booktitle={Proceedings of the 43rd International ACM SIGIR conference on research and development in Information Retrieval},
  pages={39--48},
  year={2020}
}

@article{lewis2020retrieval,
  title={Retrieval-augmented generation for knowledge-intensive nlp tasks},
  author={Lewis, Patrick and Perez, Ethan and Piktus, Aleksandra and Petroni, Fabio and Karpukhin, Vladimir and Goyal, Naman and K{\"u}ttler, Heinrich and Lewis, Mike and Yih, Wen-tau and Rockt{\"a}schel, Tim and others},
  journal={Advances in neural information processing systems},
  volume={33},
  pages={9459--9474},
  year={2020}
}

@inproceedings{chen2026beyond,
    title = "Beyond Chunking: Discourse-Aware Hierarchical Retrieval for Long Document Question Answering",
    author = "Chen, Huiyao  and
      Yang, Yi  and
      Li, Yinghui  and
      Zhang, Meishan  and
      Hu, Baotian  and
      Zhang, Min",
    editor = "Liakata, Maria  and
      Moreira, Viviane P.  and
      Zhang, Jiajun  and
      Jurgens, David",
    booktitle = "Proceedings of the 64th Annual Meeting of the {A}ssociation for {C}omputational {L}inguistics (Volume 1: Long Papers)",
    month = jul,
    year = "2026",
    address = "San Diego, California, United States",
    publisher = "Association for Computational Linguistics",
    url = "https://aclanthology.org/2026.acl-long.829/",
    doi = "10.18653/v1/2026.acl-long.829",
    pages = "18176--18198",
    ISBN = "979-8-89176-390-6",
}

@book{manning2008introduction,
  title={Introduction to information retrieval},
  author={Manning, Christopher D},
  year={2008},
  publisher={Syngress Publishing,}
}

@article{liu2024lost,
    title = "Lost in the Middle: How Language Models Use Long Contexts",
    author = "Liu, Nelson F.  and
      Lin, Kevin  and
      Hewitt, John  and
      Paranjape, Ashwin  and
      Bevilacqua, Michele  and
      Petroni, Fabio  and
      Liang, Percy",
    journal = "Transactions of the Association for Computational Linguistics",
    volume = "12",
    year = "2024",
    address = "Cambridge, MA",
    publisher = "MIT Press",
    url = "https://aclanthology.org/2024.tacl-1.9/",
    doi = "10.1162/tacl_a_00638",
    pages = "157--173",
}

@article{malkov2018efficient,
  title={Efficient and robust approximate nearest neighbor search using hierarchical navigable small world graphs},
  author={Malkov, Yu A and Yashunin, Dmitry A},
  journal={IEEE transactions on pattern analysis and machine intelligence},
  volume={42},
  number={4},
  pages={824--836},
  year={2018},
  publisher={IEEE}
}

@article{nogueira2019passage,
  title={Passage Re-ranking with BERT},
  author={Nogueira, Rodrigo and Cho, Kyunghyun},
  journal={arXiv preprint arXiv:1901.04085},
  year={2019}
}

@article{nogueira2019document,
  title={Document expansion by query prediction},
  author={Nogueira, Rodrigo and Yang, Wei and Lin, Jimmy and Cho, Kyunghyun},
  journal={arXiv preprint arXiv:1904.08375},
  year={2019}
}

@article{dong2024multi,
  title={Multi-view content-aware indexing for long document retrieval},
  author={Dong, Kuicai and Deik, Derrick Goh Xin and Lee, Yi Quan and Zhang, Hao and Li, Xiangyang and Zhang, Cong and Liu, Yong},
  journal={arXiv preprint arXiv:2404.15103},
  year={2024}
}

@inproceedings{reimers2019sentence,
  title={Sentence-bert: Sentence embeddings using siamese bert-networks},
  author={Reimers, Nils and Gurevych, Iryna},
  booktitle={Proceedings of the 2019 conference on empirical methods in natural language processing and the 9th international joint conference on natural language processing (EMNLP-IJCNLP)},
  pages={3982--3992},
  year={2019}
}

@article{robertson2004understanding,
  title={Understanding inverse document frequency: on theoretical arguments for IDF},
  author={Robertson, Stephen},
  journal={Journal of documentation},
  volume={60},
  pages={503--520},
  year={2004},
  publisher={Emerald Group Publishing Limited}
}

@article{robertson2009probabilistic,
author = {Robertson, Stephen and Zaragoza, Hugo},
title = {The Probabilistic Relevance Framework: BM25 and Beyond},
year = {2009},
issue_date = {April 2009},
publisher = {Now Publishers Inc.},
address = {Hanover, MA, USA},
volume = {3},
number = {4},
issn = {1554-0669},
url = {https://doi.org/10.1561/1500000019},
doi = {10.1561/1500000019},
journal = {Found. Trends Inf. Retr.},
month = apr,
pages = {333–389},
numpages = {57}
}

@article{salton1988termweighting,
  title={Term-weighting approaches in automatic text retrieval},
  author={Salton, Gerard and Buckley, Christopher},
  journal={Information processing \& management},
  volume={24},
  number={5},
  pages={513--523},
  year={1988},
  publisher={Elsevier}
}

@inproceedings{sarthi2024raptor,
  title={Raptor: Recursive abstractive processing for tree-organized retrieval},
  author={Sarthi, Parth and Abdullah, Salman and Tuli, Aditi and Khanna, Shubh and Goldie, Anna and Manning, Christopher},
  booktitle={International Conference on Learning Representations},
  volume={2024},
  pages={32628--32649},
  year={2024}
}

@article{singh2026beyond,
  title={Beyond the Reranker: Do RAG Retrieval Enhancements Help Once a Strong Reranker Is Present?},
  author={Singh, Sadanand and Reddy, Allam and Chopra, Manan},
  journal={arXiv preprint arXiv:2606.28367},
  year={2026}
}

@inproceedings{su2025bright,
  title={{BRIGHT}: A Realistic and Challenging Benchmark for Reasoning-Intensive Retrieval},
  author={Hongjin SU and Howard Yen and Mengzhou Xia and Weijia Shi and Niklas Muennighoff and Han-yu Wang and Liu Haisu and Quan Shi and Zachary S Siegel and Michael Tang and Ruoxi Sun and Jinsung Yoon and Sercan O Arik and Danqi Chen and Tao Yu},
  booktitle={The Thirteenth International Conference on Learning Representations},
  year={2025},
  url={https://openreview.net/forum?id=ykuc5q381b}
}

@inproceedings{sun2023chatgpt,
  title={Is ChatGPT good at search? investigating large language models as re-ranking agents},
  author={Sun, Weiwei and Yan, Lingyong and Ma, Xinyu and Wang, Shuaiqiang and Ren, Pengjie and Chen, Zhumin and Yin, Dawei and Ren, Zhaochun},
  booktitle={Proceedings of the 2023 conference on empirical methods in natural language processing},
  pages={14918--14937},
  year={2023}
}

@inproceedings{thakur2021beir,
    title={{BEIR}: A Heterogeneous Benchmark for Zero-shot Evaluation of Information Retrieval Models},
    author={Nandan Thakur and Nils Reimers and Andreas R{\"u}ckl{\'e} and Abhishek Srivastava and Iryna Gurevych},
    booktitle={Thirty-fifth Conference on Neural Information Processing Systems Datasets and Benchmarks Track (Round 2)},
    year={2021},
    url={https://openreview.net/forum?id=wCu6T5xFjeJ}
}

@inproceedings{tombros1998advantages,
author = {Tombros, Anastasios and Sanderson, Mark},
title = {Advantages of query biased summaries in information retrieval},
year = {1998},
isbn = {1581130155},
publisher = {Association for Computing Machinery},
address = {New York, NY, USA},
url = {https://doi.org/10.1145/290941.290947},
doi = {10.1145/290941.290947},
booktitle = {Proceedings of the 21st Annual International ACM SIGIR Conference on Research and Development in Information Retrieval},
pages = {2–10},
numpages = {9},
location = {Melbourne, Australia},
series = {SIGIR '98}
}

@inproceedings{wang2026paperregister,
  author = {Wang, S. and Zhang, H. and Li, J. and Liu, Y.},
  title = {PaperRegister: Hierarchical register indexing for academic document retrieval},
  booktitle = {Proceedings of the 64th Annual Meeting of the Association for Computational Linguistics},
  year = {2026},
}

@inproceedings{wilkinson1994effective,
  title={Effective retrieval of structured documents},
  author={Wilkinson, Ross},
  booktitle={SIGIR’94: Proceedings of the Seventeenth Annual International ACM-SIGIR Conference on Research and Development in Information Retrieval, organised by Dublin City University},
  pages={311--317},
  year={1994},
  organization={Springer}
}

@inproceedings{xiong2021approximate,
  author = {Xiong, L. and Xiong, C. and Li, Y. and Tang, K.-F. and Liu, J. and Bennett, P. and Ahmed, J. and Overwijk, A.},
  title = {Approximate Nearest Neighbor Negative Contrastive Learning for Dense Text Retrieval},
  booktitle = {Proceedings of the 9th International Conference on Learning Representations ({ICLR})},
  year = {2021},
}

@inproceedings{zhang2026leanrag,
  title={Leanrag: Knowledge-graph-based generation with semantic aggregation and hierarchical retrieval},
  author={Zhang, Yaoze and Wu, Rong and Cai, Pinlong and Wang, Xiaoman and Yan, Guohang and Mao, Song and Wang, Ding and Shi, Botian},
  booktitle={Proceedings of the AAAI Conference on Artificial Intelligence},
  volume={40},
  number={41},
  pages={34862--34869},
  year={2026}
}

@article{bhat2025rethinking,
  title={Rethinking chunk size for long-document retrieval: A multi-dataset analysis},
  author={Bhat, Sinchana Ramakanth and Rudat, Max and Spiekermann, Jannis and Flores-Herr, Nicolas},
  journal={arXiv preprint arXiv:2505.21700},
  year={2025}
}

@inproceedings{reuter2025towards,
    title = "Towards Reliable Retrieval in {RAG} Systems for Large Legal Datasets",
    author = "Reuter, Markus  and
      Lingenberg, Tobias  and
      Liepina, Ruta  and
      Lagioia, Francesca  and
      Lippi, Marco  and
      Sartor, Giovanni  and
      Passerini, Andrea  and
      Sayin, Burcu",
    editor = "Aletras, Nikolaos  and
      Chalkidis, Ilias  and
      Barrett, Leslie  and
      Goanț{\u{a}}, C{\u{a}}t{\u{a}}lina  and
      Preoțiuc-Pietro, Daniel  and
      Spanakis, Gerasimos",
    booktitle = "Proceedings of the Natural Legal Language Processing Workshop 2025",
    month = nov,
    year = "2025",
    address = "Suzhou, China",
    publisher = "Association for Computational Linguistics",
    url = "https://aclanthology.org/2025.nllp-1.3/",
    doi = "10.18653/v1/2025.nllp-1.3",
    pages = "17--30",
    ISBN = "979-8-89176-338-8",
}
\end{document}